\documentclass[aps,prl,reprint,preprintnumbers,superscriptaddress,nofootinbib,longbibliography,floatfix]{revtex4-2}
\pdfoutput=1
\usepackage{rotating}
\usepackage{array}
\usepackage{amsmath}
\usepackage[normalem]{ulem}
\usepackage{slashed}
\usepackage{booktabs}
\usepackage[pdftex,table]{xcolor}
\usepackage{units}
\usepackage{xfrac}
\usepackage{mathtools}
\usepackage{empheq}
\usepackage[]{units}
\usepackage{multirow}
\usepackage{amssymb}
\usepackage{url}
\usepackage{comment}
\usepackage{physics}
\usepackage{color,soul}
\usepackage{bbm}
\usepackage[caption=false]{subfig}
\usepackage{adjustbox}
\usepackage[T1]{fontenc}

\usepackage{hyperref}
\hypersetup{
  colorlinks=true,
  citecolor=blue,
  linkcolor=blue,
  urlcolor=blue
}

\newcommand{\pt}{\mathrm{p_{T}}}

\begin{document}

\title{Solving Key Challenges in Collider Physics with Foundation Models}

\author{Vinicius Mikuni}
\email{vmikuni@lbl.gov}
\affiliation{National Energy Research Scientific Computing Center, Berkeley Lab, Berkeley, CA 94720, USA}

\author{Benjamin Nachman}
\email{bpnachman@lbl.gov}
\affiliation{Physics Division, Lawrence Berkeley National Laboratory, Berkeley, CA 94720, USA}
\affiliation{Berkeley Institute for Data Science, University of California, Berkeley, CA 94720, USA}

\begin{abstract}
    Foundation Models are neural networks that are capable of simultaneously solving many problems.  Large Language Foundation Models like ChatGPT have revolutionized many aspects of daily life, but their impact for science is not yet clear.  In this paper, we use a new Foundation Model for hadronic jets to solve three key challenges in collider physics.  In particular, we show how experiments can (1) save significant computing power when developing reconstruction algorithms, (2) perform a complete uncertainty quantification for high-dimensional measurements, and (3) search for new physics with model agnostic methods using low-level inputs.  In each case, there are significant computational or methodological challenges with current methods that limit the science potential of deep learning algorithms.  By solving each problem, we take jet Foundation Models beyond proof-of-principle studies and into the toolkit of practitioners.  
\end{abstract}

\maketitle

Deep learning is enabling researchers to analyze high-dimensional data holistically without first projecting into low-dimensional summary statistics.  New methods in areas such as parameter estimation~\cite{Cranmer:2019eaq}, deconvolution~\cite{Arratia:2021otl,Du:2024gbp}, and anomaly detection~\cite{Kasieczka:2021xcg,Balazs:2021uhg,2112.03769} are enabling data analyses that were unimaginable before deep learning.  We are also starting to see the first results with real data~\cite{collaboration2020dijet,ATLAS:2023ixc,ATLAS:2023azi,CMS-PAS-EXO-22-026,H1:2021wkz,H1prelim-22-031,H1:2023fzk,H1prelim-21-031,LHCb:2022rky,Komiske:2022vxg,Song:2023sxb}.  While the machine learning methods and computing exist to train these methods, there are still critical challenges to deploying them in practice and scaling them to the full phase space.  The state-of-the-art machine learning architectures are data hungry and require (and can leverage~\cite{Batson:2023ohn}) massive datasets for training.  In many cases, we simply cannot generate datasets large enough - either because the full simulation of a detector is too slow or because running an experiment longer is too expensive.  Without new meta methods, we will not be able to harness the power of modern machine learning to make the most of our precious data.

Foundation models have emerged from industry as meta methods capable of enabling a diverse set of downstream deep learning tasks.   These models are trained with enormous and diverse datasets that can then be fined-tuned to other use-cases with little or no data.  Most of the excitement around foundation models has been in the area of text, where Large Language Models (LLMs) have clearly demonstrated paradigm-shifting capabilities for the way we interact with the world around us.   LLMs are trained on large corpuses of text to fill in the blanks from masked out words or sentences.  There have been some proposals in particle physics to use this and other types of \textit{self-supervised representation learning} for datasets composed of particles~\cite{Dillon:2021gag,Dillon:2022tmm,Dillon:2023zac,Heinrich:2024sbg,Harris:2024sra,Kuh:2024lgx}.   An alternative approach is to promote transfer learning to a foundation model through \textit{supervised representation learning}.  In this case, a model is trained on one dataset and then applied elsewhere.  A number of particle physics studies have used transfer learning of this kind~\cite{Kuchera:2018djs,Chappell:2022yxd,Dreyer:2022yom,Beauchesne:2023vie}, but only consider one downstream task and so are not yet foundation models.  Across all of these promising studies, there has not yet been a demonstrated foundation model that has been used to solve open scientific challenges.

Our goal is to show that foundation models for particle physics can be used to address outstanding challenges with realistic datasets.  We make use of a new foundation model called \textsc{OmniLearn} based on supervised representation discussed in detail in Ref.~\cite{prd}.  \textsc{OmniLearn} was trained on a large dataset and can provide state-of-the-art results for accuracy, precision, and/or speed of many downstream tasks
involving hadronic final states across detectors and collision systems and for both classification and generation. Interestingly, by aligning the training with the inference goals, \textsc{OmniLearn} is a compact model with less than 2M parameters. This model size fits comfortably in a single Graphical Processing Unit (GPU) and empowers scientists with restricted computing resources to benefit.

Despite its small size, we consider \textsc{OmniLearn} to be a foundation model, that can address some of the most important challenges in machine learning for collider
physics. Here, we show results for the most promising applications starting with a simulation challenge. Full detector simulations at the LHC are computationally expensive and generating large enough datasets to train state-of-the-art models is becoming a limiting factor for many analyses.  \textsc{OmniLearn} was trained using data from a public fast simulation - can such a model reduce the number of fully simulated events required to achieve cutting edge performance?  Second, we address a training challenge.  A growing number of machine learning-based methods require the estimation of likelihood ratios.  Uncertainties on this estimate require re-running the learning thousands or tens of thousands of times~\cite{H1:2021wkz,H1:2023fzk}, which is prohibitive for full phase-space inference.  Can \textsc{OmniLearn} reduce the time required to train these models so that the full uncertainty quantification can be achieved?  Third, we address an anomaly detection challenge.  Anomaly detection methods that are trained directly on data are fundamentally limited by the training dataset size.  As a result, high-dimensional methods struggle to obtain the relevant sensitivity to rare signals in high-dimensional feature spaces~\cite{Buhmann:2023acn}.  Can \textsc{OmniLearn} push the sensitivity of such methods to find signals that would not have been found before?

The details of \textsc{OmniLearn} can be found in Ref.~\cite{prd}, but are briefly summarized in the following. The backbone network of \textsc{OmniLearn} leverages modern developments for jet representation, using a combination of attention mechanisms~\cite{DBLP:journals/corr/VaswaniSPUJGKP17} and dynamic convolutional operations~\cite{DBLP:journals/corr/abs-1801-07829} to improve both the global and local description of particles clustered inside jets.  This model is named a Point-Edge Transformer (\textsc{PET}), consisting of a shared representation (\textsc{PET} body) and two task specific heads, used for the tasks of classification and particle generation. The introduction of tasks specific network components makes the model modular, and able to discard irrelevant heads during downstream tasks, thus reducing even further the model size during adaptation. The pre-trained PET model is then referred to as \textsc{OmniLearn}. \textsc{OmniLearn} is trained using the JetClass dataset~\cite{Qu:2022mxj}, consisting of 100M jets and featuring 10 different jet classes.  Each jet is represented as a set of constituent particles.

Jets are ubiquitous in high-energy collider physics and identifying the origin of a jet (`jet tagging') is a key component of a multitude of direct searches for new phenomena and precision measurements.  Improvements in jet tagging performance directly translate into improved sensitivity of many analyses.  State-of-the-art jet tagging models require many tens of millions of jets for training.  As full detector simulations are expensive, it is prohibitive to generate large enough datasets for every tagging task and for every time there are changes in the simulation set (e.g. better particle-level description or different operational conditions).  Our hypothesis is that we can adapt a foundation model trained on less accurate fast simulation with a small sample of realistic simulations to achieve competitive performance on the full realistic problem.

We exemplify this idea showing the results obtained by \textsc{OmniLearn} adapted to the publicly available ATLAS Top tagging dataset~\cite{ATL-PHYS-PUB-2022-039}. In this dataset, 40M events are generated with \textsc{Pythia8} using the NNPDF2.3LO~\cite{Ball:2012cx} set of parton distribution functions and the A14~\cite{Buckley:2014ctn} set of tuned parameters. Pileup effects are simulated by overlaying inelastic interactions on top of the underlying hard scattering process based on the 2017 data taking period. Hadronic boosted top quarks are obtained in simulated events containing the decay of a heavy $Z'$ boson with mass of 2 TeV. Background jets are obtained from simulations of generic dijet events. Unified Flow Objects~\cite{ATLAS:2020gwe} are used to determine the jet constituents. Jets are clustered using anti-$k_{T}$ algorithm~\cite{Cacciari:2005hq,Cacciari:2011ma,Cacciari:2008gp} with R=1.0 with additional pileup mitigation algorithms~\cite{Berta:2014eza,Berta:2019hnj,Cacciari:2014gra} applied. The Soft-Drop algorithm~\cite{Larkoski:2014wba} is also applied to remove soft and wide-angle radiation. Note that unlike the JetClass dataset, the ATLAS top tagging dataset features full detector simulation, event reconstruction, and pileup particles. 

We consider the scenario where we use the entire training data for the adaptation of \textsc{OmniLearn} and when only 10\% of the data is used (4M events). Results are reported in Table.~\ref{tab:results_atlas} for four metrics: the area under curve (AUC), accuracy for a fixed threshold of 0.5, and the inverse background efficiency at two fixed values of the signal efficiency. We observe that \textsc{OmniLearn} excels the performance of all previously reported benchmarks in this dataset and is able to match the previously best performing model using only 10\% of the data, thus requiring significantly less examples to achieve state-of-the-art performance. 

\begin{table}[th]
    \centering
    \caption{Comparison between the performance reported for different classification algorithms on the ATLAS top tagging dataset. Bold results represent the algorithm with highest performance. All results besides \textsc{OmniLearn} and \textsc{PET} are taken from~\cite{ATL-PHYS-PUB-2022-039}. Five \textsc{PET} and \textsc{OmniLearn} models are trained and the error of multiple predictions is at the last decimal place and hence not shown.}
    \label{tab:results_atlas}
	\begin{tabular}{lccccc}
    \noalign{\smallskip}\hline
          &  AUC &Acc & \multicolumn{2}{c}{1/$\epsilon_B$} \\
          \cline{4-5}
          & & & $\epsilon_S = 0.5$ & $\epsilon_S = 0.8$ \\
            \hline
            ResNet 50 & 0.885 & 0.803 & 21.4 & 5.13 \\
            EFN & 0.901 & 0.819 & 26.6 & 6.12 \\
            hlDNN & 0.938 & 0.863 & 51.5 & 10.5 \\
            DNN & 0.942 & 0.868 & 67.7 & 12.0  \\
            PFN & 0.954 &  0.882 & 108.0 & 15.9  \\
            ParticleNet & 0.961 & 0.894 & 153.7 & 20.4 \\
            \hline
            \textsc{PET} classifier (4M) & 0.959&0.890 &146.5 & 19.4\\
            \textsc{OmniLearn} (4M) & 0.961 & 0.894 &172.1  & 20.8\\
            \textsc{PET} classifier (40M) & 0.964 & 0.898& 201.4 & 23.6\\
            \textsc{OmniLearn} (40M) & \textbf{0.965}&\textbf{0.899} & \textbf{207.3} & \textbf{24.1}\\
	\noalign{\smallskip}
	\end{tabular}
\end{table}

Correcting physics measurements for detector distortions enables efficient comparisons between measurements and theory predictions. This technique is known in collider physics as \textit{unfolding}. Machine learning greatly increase the flexibility and potential of unfolding by allowing the simultaneous correction of multiple distributions without the use of histograms~\cite{Arratia:2021otl}. The \textsc{OmniFold} algorithm~\cite{Andreassen:2019cjw,Andreassen:2021zzk} introduced an iterative approach for unfolding based on learned classifiers that use the data collected by experiments to determine the unfolded distributions. Since the methodology relies on the data availability, physics processes with limited data can severely restrict the precision the algorithm can achieve. In contrast, the general representation of jets learned by \textsc{OmniLearn} can compensate for the limited data. We verify this observation adapting \textsc{OmniLearn} for complete unfolding using all available features in the dataset introduced in Ref.~\cite{Andreassen:2019cjw}. The dataset consists of proton-proton collisions producing a $Z$ boson, generated at a center-of-mass energy of $\sqrt{s}=14$ TeV. A sample used as the `data' representative is simulated using particle collisions with the default tune of Herwig 7.1.5~\cite{Bahr:2008pv,Bellm:2015jjp,Bellm:2017bvx}. A second dataset, representative of the `simulation' we want to correct, is simulated using \textsc{Pythia8} with Tune 21~\cite{ATL-PHYS-PUB-2014-021}. Detector distortions are simulated with \textsc{Delphes} and the CMS tune that uses a particle flow reconstruction. Jets are clustered using  the anti-$k_T$ algorithm with radius parameter $R=0.4$ as implemented in FastJet~3.3.2 package~\cite{Cacciari:2011ma}. The $Z$ bosons are required to have $\pt>200$ GeV in order to mitigate acceptance effects.  Note that the corresponding jets are much lower momentum than those in JetClass (between 500-1000 GeV).

\begin{figure}[ht]
    \centering
        \includegraphics[width=.4\textwidth]{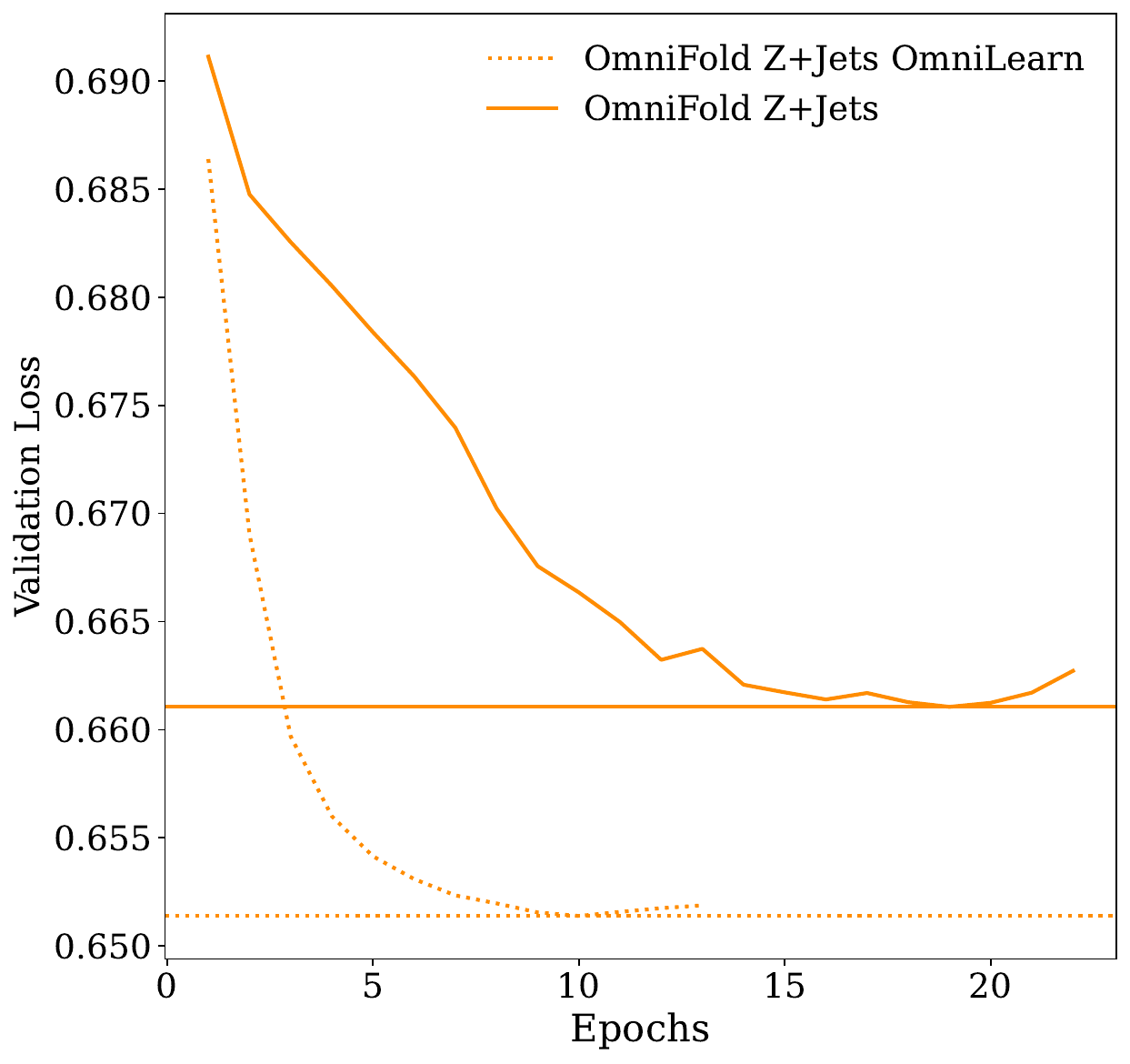}
    \caption{Validation loss curves obtained in the first iteration and first step of the \textsc{OmniFold} dataset. The \textsc{OmniLearn} validation loss is compared with the \textsc{PET} classifier trained from scratch.}
    \label{fig:loss_omnifold}
\end{figure}

A key challenge with \textsc{OmniFold} is that it requires training $2n$ neural networks every time the algorithm is run, typically with $n=4$ iterations.  When ensembling to improve precision and bootstrapping to estimate uncertainties, this quickly adds up to tens of thousands of networks that must be trained.  In Figure~\ref{fig:loss_omnifold}, we show the validation loss obtained in an \textsc{OmniFold} step using \textsc{OmniLearn} and starting the classification task from scratch.  \textsc{OmniLearn} is able to converge twice as fast and reach a lower minimum comparing to training from scratch. The overall speed up time from \textsc{OmniLearn} is also close to a factor 2 since the most time-consuming iteration is the first one with consecutive iterations starting from the previous iteration checkpoint and requiring few epochs for convergence.  This lower validation loss directly translates into better performance on physics metrics, calculated as the triangular distance between different physics observables introduced in~\cite{Andreassen:2019cjw} and summarized in Table~\ref{tab:omnifold}.  The unfolding is performed at the level of all particle momenta and yet \textsc{OmniLearn} outperforms the classical \textsc{OmniFold} using all particles with a \textsc{DeepSets} architecture which outperforms the classical bin-based, dedicated unfolding (IBU~\cite{DAgostini:1994fjx}).

\begin{table}[ht]
    \centering
	\footnotesize
    \caption{Comparison of the triangular discriminator~\cite{850703,Gras:2017jty,Bright-Thonney:2018mxq} between different algorithms for unfolding. Uncertainties from \textsc{PET} and \textsc{OmniLearn} are taken from 100 histogram variations within the statistical uncertainty of the prediction. Quantities in bold represent the method with best performance. IBU and \textsc{OmniFold} results with \textsc{DeepSets} is taken from~\cite{Andreassen:2019cjw}}
    \label{tab:omnifold}
    \begin{tabular}{lcccc}
        \hline
        Metric & IBU & \multicolumn{3}{c}{\textsc{OmniFold}} \\
        \cline{3-5}
        & &  DeepSets & \textsc{PET} classifier & \textsc{OmniLearn} \\
        \hline
        Jet mass  & 9.31 & \textbf{2.77} & \textbf{2.8$\pm$0.9} & \textbf{2.6$\pm$0.8} \\
        N  & 1.51 & \textbf{0.33} & 0.50$\pm$0.15 & \textbf{0.34$\pm$0.1} \\
        Jet Width  & 0.11 & 0.10 & 0.09$\pm$0.02 & \textbf{0.07$\pm$0.01} \\
        $\log\rho$ & 0.71 & 0.35 & 0.23$\pm$0.07 & \textbf{0.14$\pm$0.03} \\
        $\tau_{21}$ & 1.10 & 0.53 & 0.13$\pm$0.03 & \textbf{0.05$\pm$0.01} \\
        $z_g$  & 0.37 & 0.68 & \textbf{0.19$\pm$0.03} & \textbf{0.21$\pm$0.04} \\
        \hline
    \end{tabular}
\end{table}

Like \textsc{OmniFold}, many machine learning-based anomaly detection protocols require training directly on data.  One particularly powerful approach with asymptotic guarantees of optimality~\cite{Metodiev:2017vrx} is to train a classifier to distinguish data from a background-only reference dataset~\cite{Collins:2019jip}.  If the reference is accurate and the classifier is powerful, any non-trivial performance could be an indication of something new.  Approaches largely differ in how they generate the reference dataset.  In the case of resonant anomalies, sideband information can be used to build the reference.  A state-of-the-art approach is CATHODE~\cite{Hallin:2021wme}, which trains a conditional generative model in the sideband, interpolates the model to the signal region, samples a reference dataset, and then trains a classifier to distinguish the synthetic data from the real data in the signal region.  While originally developed for high-level features, CATHODE was recently studied for processing entire jets~\cite{Buhmann:2023acn}.  While it was able to discover signals injected into a background-only dataset, the amount of signal had to be so high that it would have also been discoverable without doing any machine learning.  

We reexamine the high-dimensional CATHODE approach using \textsc{OmniLearn}.  For this purpose, we use the R\&D dataset from the LHC Olympics data challenge~\cite{LHCOlympics,Kasieczka:2021xcg}.  The background consists of generic dijet events while the signal is a resonant boson production $A \to B (\to q q') C(\to q q') $ with masses $m_{A}, m_B, m_C$ = 3.5, 0.5, 0.1~TeV, respectively.  Signal and background events are generated with \textsc{Pythia8} interfaced with \textsc{Delphes3.4.1} for detector simulation. Jets are defined using the anti-$k_T$ algorithm as implemented in \textsc{FastJet} with $R=1$. We focus on the two leading jets in transverse momentum space and require the leading jet to have $\pt > 1.2$~TeV. After selection, we save all particles associated to the two most energetic jets, resulting in a maximum particle multiplicity of 279 particles per jet. We need to use \textsc{OmniLearn} twice: once to generate the background and once to distinguish the resulting synthetic samples from the data.  Since there are two jets, we use a \textsc{DeepSets}~\cite{deepsets} to mix the two jets into a full event model~\cite{prd}.

\begin{figure}[ht]
    \centering
        \includegraphics[width=.45\textwidth]{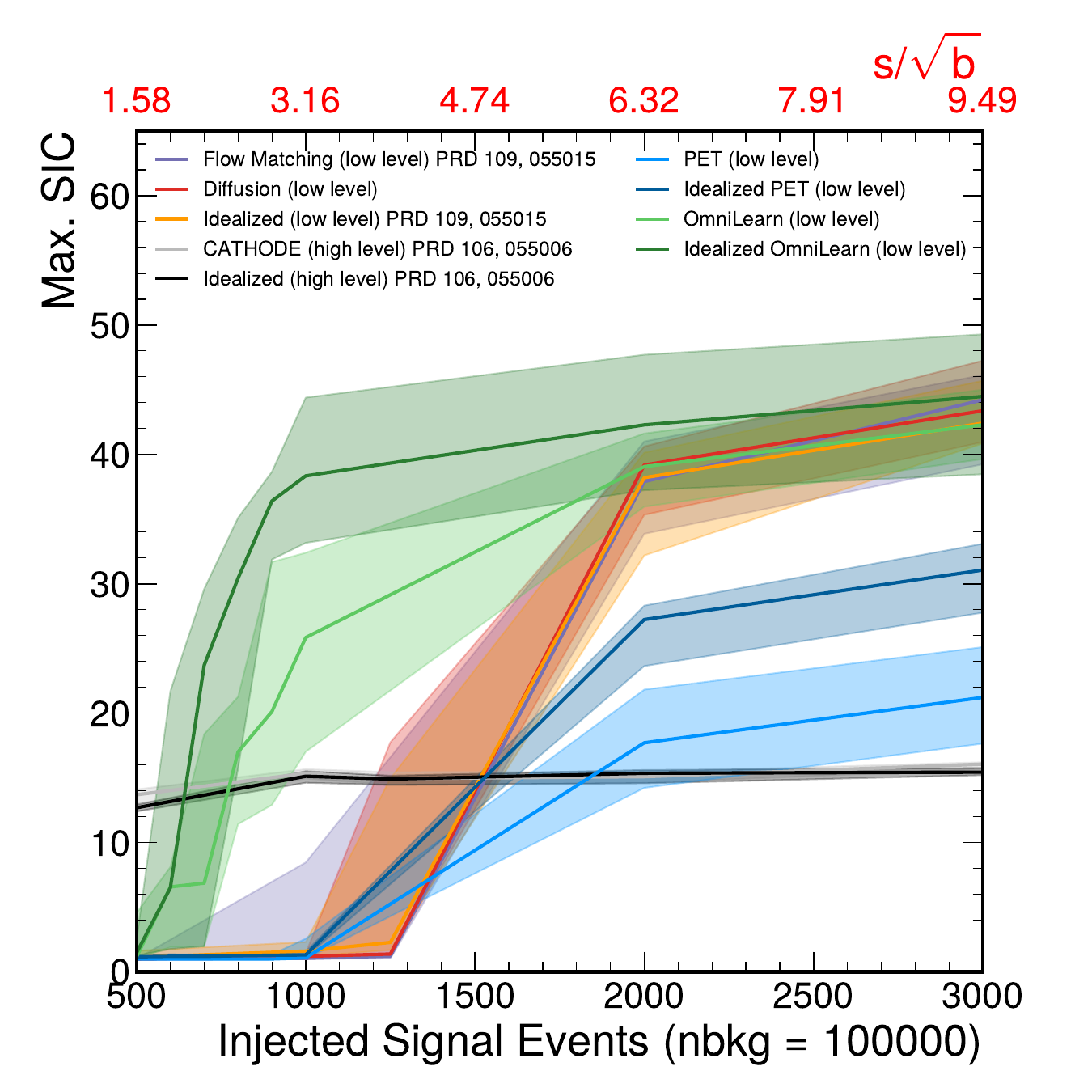}
    \caption{Maximum values of the SIC curve evaluated over different values of injected signal. \textsc{OmniLearn} and \textsc{PET} classifier results are compared with other algorithms used for the same task.}
    \label{fig:max_sic}
\end{figure}

The performance of \textsc{OmniLearn} on the high-dimensional anomaly detection task is shown in Fig.~\ref{fig:max_sic}.  Our figure of merit is the maximum significance improvement characteristic curve (SIC).  This is the multiplicative factor by which the significance improves with an ideal selection on the classifier.  Values above unity indicate value added and when multiplied by the upper, horizontal axis, the combined number indicates the final significance.  The results are presented for \textsc{OmniLearn}, \textsc{PET}, the simpler model of Ref.~\cite{Buhmann:2023acn}, and an idealized scenario where the background is a statistically identical sample of true background events in the signal region (no learning).   \textsc{OmniLearn} shows non-negligible signal sensitivity for signals injections above 600 events, corresponding to an initial significance $S/\sqrt{B}\sim 2$, representing a large improvement in sensitivity compared to previous results that required signal injections above 1400 ($S/\sqrt{B}\sim 4$). Conversely, training the model from scratch shows lower performance compared to results reported in~\cite{Buhmann:2023acn}. The reason for this difference is due to the limited amount of data in the signal region (around 100k). Even when using a larger generated background of 350k events, the dataset size of this application is at least 2 times smaller than all previous datasets investigated so far. 

In summary, we have demonstrated that our pre-trained transformer approach~\cite{prd} can
be considered as a foundation model for jet physics and is ready to join the toolkit of particle physicists. This is demonstrated using the \textsc{OmniLearn} foundation model~\cite{prd} in the context of three potentially transformative applications:

\begin{enumerate}
    \item For tasks with \textbf{expensive simulations}, we have shown that \textsc{OmniLearn} can effectively amplify the training statistics to improve the performance.  For example, on a full detector simulation dataset for top tagging from ATLAS~\cite{ATL-PHYS-PUB-2022-039}, we show that we can achieve (or exceed) state-of-the-art performance with only 10\% of the training dataset.  This could lead to significant computational savings for the experiments when developing new taggers.  It opens up the possibility of building analysis-specific taggers without generating large training datasets for every analysis.
    \item For full phase space \textbf{unfolding}, \textsc{OmniLearn} is not only more precise, but also much faster - converging in about half the time as a dedicated approach.  This is critical for practical applications of unfolding, where ensembling and statistical uncertainties require the training of a computationally expensive number of models~\cite{H1:2021wkz,H1prelim-22-031,H1:2023fzk,H1prelim-21-031,LHCb:2022rky,Song:2023sxb}.   By improving the accuracy, it also will enhance the diversity of downstream analysis tasks that can benefit from high-dimensional, unbinned data.
    \item For \textbf{anomaly detection}, we have shown for the first time that full phase space methods are capable of non-trivial discoveries.  Previous work in resonant anomaly detection~\cite{Buhmann:2023acn} had shown that full phase-space methods could significantly amplify injected signals, but only if their starting significance was well above 2.  With \textsc{OmniLearn}, this is pushed down to 2.
\end{enumerate}

We are excited to explore (many) other applications of such approaches; all of the code and data to reproduce these results are also public to facilitate other researchers to do the same.  We also envision a number of extensions, including the addition of even more training data for the foundation model.  The way we train machine learning models has forever changed - instead of starting from scratch, it will be more effective to start with a foundation model.  If the data are sufficiently different that \textsc{OmniLearn} is not applicable, then the tools used to train it can be readily adapted to create a library of foundation models to enable all tasks in particle physics and beyond.

\section*{Code Availability}

The code for this paper can be found at \url{https://github.com/ViniciusMikuni/OmniLearn}.

\section*{Acknowledgments}
We thank F. Dreyer, Ming Fong, R. Grabarczyk, K. Greif, P.F. Monni, and D. Whiteson for interesting discussions related to transfer learning.  We also thank J. Birk,  A. Hallin , and G. Kasieczka for comments on the manuscript
Additionally, we thank our colleagues from the H1 Collaboration for allowing us to use the simulated MC event samples. We also thank DESY-IT and the MPI f\"ur Physik for providing computing infrastructure and supporting the data preservation project of the HERA experiments.
VM, and BN are supported by the U.S. Department of Energy (DOE), Office of Science under contract DE-AC02-05CH11231.  This research used resources of the National Energy Research Scientific Computing Center, a DOE Office of Science User Facility supported by the Office of Science of the U.S. Department of Energy under Contract No. DE-AC02-05CH11231 using NERSC awards HEP-ERCAP0021099 and HEP-ERCAP0028249.

\bibliography{HEPML,other}
\bibliographystyle{apsrev4-1}

\end{document}